\documentclass[a4paper]{article}

\usepackage{INTERSPEECH2022}
\usepackage{hyperref}

\title{Adapting TTS models For New Speakers using Transfer Learning}
\name{*Paarth Neekhara$^1$\thanks{* Work performed as an intern at NVIDIA} \qquad Jason Li$^{2}$ \qquad Boris Ginsburg$^{2}$}
\address{$^{1}$ University of California, San Diego \\
      {$^{2}$}NVIDIA, Santa Clara }
\email{pneekhar@eng.ucsd.edu, jasoli@nvidia.com, bginsburg@nvidia.com}

\begin{document}

\maketitle
\begin{abstract}
  Training neural text-to-speech (TTS) models for a new speaker typically requires several hours of high quality speech data. Prior works on voice cloning attempt to address this challenge by adapting pre-trained multi-speaker TTS models for a new voice, using a few minutes of speech data of the new speaker. However, publicly available large multi-speaker datasets are often noisy, thereby resulting in TTS models that are not suitable for use in products. 
  We address this challenge by proposing transfer-learning guidelines for adapting high quality single-speaker TTS models for a new speaker, using only a few minutes of speech data. We propose two finetuning methods that significantly reduce the data requirement and training time for adapting pre-trained spectrogram synthesizers and vocoders for a new voice.
  We conduct an extensive study using different amounts of data for a new speaker and evaluate the synthesized speech in terms of naturalness and voice/style similarity to the target speaker. We find that fine-tuning a single-speaker TTS model on just 30 minutes of data, can yield comparable performance to a model trained from scratch on more than 27 hours of data for both male and female target speakers.
\end{abstract}

\section{Introduction}
\label{sec:intro}

Training high-quality text-to-speech (TTS) models is desirable for several applications such as natural sounding voice assistants, voice-overs in animated films and computer generated advertisement campaigns. Typically, TTS models are trained on 10-20 hours of transcribed speech data for a given speaker. While such models have shown great promise in generating natural sounding speech, the process of generating a new synthetic voice is very costly and time-consuming. The process requires recording a new speaker's voice, cleaning up the recorded data and training the spectrogram synthesis and vocoder models. Since many speaking characteristics for a given language are shared amongst different speakers, it is desirable to reuse the knowledge of a pre-trained TTS model when learning to synthesize the voice of a new speaker.

Prior works~\cite{neuralvoicecloning,transferspeakerverification} have utilized transfer learning techniques to adapt pre-trained TTS models for a new speaker. In these works, a multi-speaker TTS model is trained on a large speaker-diverse dataset containing several hours of audio from hundreds or thousands of speakers.  In some of these works, the speaker embedding layer is replaced by a pre-trained speaker encoder. This encoder was trained for the task of speaker verification and is used to derive the speaker embedding for a new speaker from a few reference audio samples. To adapt the pre-trained model for a new speaker, these models can either be conditioned directly on the derived speaker embedding of the new speaker (zero-shot voice cloning), or they can be fine-tuned on a few minutes of transcribed speech samples of the new speaker using transfer learning. 

While the above works show promise in generating synthetic speech for a new speaker, the results fall short in terms of naturalness and audio quality. This is because most large-scale public datasets for TTS are usually noisy and cannot be used to train production-level models. Using multi-speaker TTS datasets for pre-training is a reasonable approach for zero-shot voice cloning, however, it is not clear whether multi-speaker TTS models offer any advantage over single-speaker TTS models when fine-tuning them for a new speaker. 

In our work, we study transfer learning techniques for adapting single-speaker TTS models for a new speaker. We use a pre-trained FastPitch~\cite{lancucki2021fastpitch} spectrogram synthesizer and a pre-trained HiFiGAN~\cite{kong2020hifi} vocoder as the base TTS system. We fine-tune this pre-trained system for a male and a female speaker using varying amounts of data ranging from one minute to an hour using two main approaches --- 1) We fine-tune the models only on the data of the new speaker, 2) We fine-tune the models on the data of both the new and the original speaker on which the base TTS system was trained. For the second approach we adapt the base FastPitch model to incorporate speaker conditioning during training. 
We find that using just 30 minutes of transcribed audio of the new speaker, our method can yield comparable results to a model trained from scratch on more than 27 hours of data of the new speaker. Our method which finetunes both FastPitch and HiFiGAN is over 15x faster than training one model (FastPitch) from scratch on a single-GPU. The main contributions of this work are as follows:
\begin{itemize}
    \item We present transfer learning methods and guidelines for finetuning single-speaker TTS models for a new voice. 
    \item We evaluate the synthesized audio in terms of speech naturalness, speaking rate, and speaker similarity with the target speaker and provide a detailed analysis of the respective metrics with varying amount of data.
    \item Our results demonstrate that transfer learning can substantially reduce the training time and amount of data needed for synthesizing a new voice. 
    \item We open-source our voice cloning framework and provide a demo to clone a new voice from a few minutes of speech data of any speaker.~\footnote{Code and Demo: \href{https://paarthneekhara.github.io/tlfortts/}{https://paarthneekhara.github.io/tlfortts/} }  
\end{itemize}

\section{Background and Related Work}

State-of-the-art neural approaches for natural TTS synthesis~\cite{ping2017deep,shen2018natural} typically decompose the waveform synthesis pipeline into two steps: (1) Synthesizing mel spectrograms from language using an attention based sequence-to-sequence model like Tacotron~\cite{tacotron, shen2018natural} or FastPitch~\cite{lancucki2021fastpitch}. (2) Vocoding the synthesized spectrograms to audible waveforms using a neural vocoder~\cite{wavenet,waveglow,advoc,kong2020hifi}. Multi-speaker TTS models~\cite{gibiansky2017deep,Ping2017DeepV3} extend this line of work by additionally conditioning the spectrogram synthesis model on speaker embeddings, which are trained end-to-end using the speaker labels in the TTS dataset. While such neural TTS models~\cite{tacotron,lancucki2021fastpitch} have shown promise in synthesizing natural sounding speech, they cannot be directly used to synthesize speech for a new speaker.

To synthesize the voice of a target speaker, past works have investigated techniques like \textit{voice conversion} and \textit{voice cloning}. The goal of voice conversion is to modify an utterance from a source speaker to make it sound like a target speaker. A commonly used technique to achieve this goal is dynamic frequency warping that aligns the spectra of different speakers. More recently, spectral conversion is performed using encoder-decoder neural networks which are typically trained on speech pairs of target and source speakers. 

Unlike voice conversion, the goal of voice cloning is to synthesize speech for a new speaker (with limited speech data) for any unseen text. 
Transfer learning has been shown to be effective in reusing the knowledge of a pre-trained machine learning model for various downstream tasks in limited data settings~\cite{seo2021wav2kws,hussain2021multi,shibano2021speech,castellon2021calm,choi2017transfer}. 
Recent works on voice cloning have leveraged transfer learning to design systems that can learn to synthesize a person’s voice from only a few audio samples~\cite{neuralvoicecloning,transferspeakerverification,chen2019cross,icaspVC,9053104,neekhara2021expressive,yourtts}. 
Many of these methods use a multi-speaker TTS model conditioned on a speaker encoder to enable zero-shot voice cloning. The speaker encoder is trained independently for the task of speaker verification on a large speaker-diverse dataset. The trained speaker encoder is then plugged into the spectrogram synthesizer as a frozen or trainable component. This joint setup is then trained on text and speech pairs from multiple speakers. To clone a new voice, the spectrogram-synthesizer can be conditioned directly on the speech samples of the new speaker to perform zero-shot voice cloning~\cite{neuralvoicecloning,transferspeakerverification}. Alternatively, the spectrogram synthesizer can also be finetuned on the text and speech pairs of the new speaker~\cite{neuralvoicecloning,neekhara2021expressive,yourtts}.

The limitation of the above voice cloning techniques is that they require a large speaker-diverse dataset during training. Since such large-scale multi-speaker datasets are usually noisy, these methods result in models that unsuitable for user-facing applications. 
In our work, we explore transfer learning methods for single-speaker TTS models and compare the results with models trained from scratch on several hours of clean speech of the target speaker.

\section{Methodology}

\subsection{Spectrogram synthesizer}
For spectrogram synthesis from text, we use FastPitch~\cite{lancucki2021fastpitch} and add a learnable alignment module~\cite{badlani2021tts} that does not require ground-truth durations. FastPitch is composed of two feed-forward transformer (FFTr) stacks, the first of which operates on the input tokens of phonemes: $x= (x_1, \cdot, x_n)$. The first FFTr outputs a hidden representation $h= FFTr(x)$ which is used to predict the duration and average pitch of every token. 
\begin{align}
\begin{split}
  & \hat{d} = \textit{DurationPredictor}(h), \\
  & \hat{p} = \textit{PitchPredictor}(h),
\end{split}
\end{align}

where $\hat{{d}}\in \mathbb{N}^n$ and $\hat{{p}}\in\mathbb{R}^n$.
Next, the pitch is projected to match the dimensionality of the hidden representation
${h}\in{R}^{n\times d}$
and added to ${h}$.
The resulting sum ${g}$ is discretely up-sampled and passed
to the second FFTr, which produces the output mel-spectrogram sequence
\begin{align}
\begin{split}
  {g} & = {h} + \text{PitchEmbedding}({p})\\
  \hat{{y}} & = \text{FFTr}(
  [
    \underbrace{g_1, \ldots, g_1}_{d_1},
    \ldots
    \underbrace{g_n, \ldots, g_n}_{d_n}
  ]).
\end{split}
\end{align}

To guide the duration prediction module, we use the learnable alignment-module and loss ($\mathcal{L}_\textit{align}$) as described in~\cite{badlani2021tts}. The alignment module combines forward-sum algorithm, the Viterbi algorithm, and a simple and efficient static prior to align the text sequence with the mel-spectrogram.
The output $d$ of this alignment module, serve as pseudo-targets for the duration prediction module. 

For guiding the pitch-prediction module, we use the ground truth $p$, derived using PYIN~\cite{de2002yin}, averaged over the input tokens using $\hat{d}$. 
The model optimizes mean-squared error between the predicted and ground-truth modalities and the forward-sum alignment loss $\mathcal{L}_\textit{align}$ as follows:
\begin{equation}
  \mathcal{L} = \lVert \hat{{y}} - {y}\rVert^2_2 + 
    \alpha      \lVert \hat{{p}} - {p}\rVert^2_2 + 
    \beta      \lVert \hat{{d}} - {d}\rVert^2_2 + 
    \gamma      \mathcal{L}_\textit{align}
\end{equation}

Our spectrogram synthesizer is trained end-to-end on text and speech pairs. Mel spectrograms $y$ and pitch $p$ are computed in the data-loading pipeline. During inference, we use the predicted $\hat{{p}}$ and $\hat{{d}}$ to synthesize speech directly from text.

\subsection{Vocoder}
For decoding the synthesized mel-spectrograms into waveforms, we use HiFiGAN~\cite{kong2020hifi} which achieves state-of-the-art results. HiFiGAN includes a generator network comprised of operating of transposed convolutions that upsample mel-spectrograms to audio.
The sum of outputs from multiple residual block is used to generate raw waveforms which are then fed as input to two discriminator networks: multi-period discriminator and multi-scale discriminator. 
The multi-period discriminator consists of small sub-discriminators, each of which obtains only specific periodic parts of raw waveform. The multi-scale discriminator consists of multiple sub-discriminators to
judge audios in different scales and thereby learn to capture consecutive patterns and long-term dependencies of the waveform. HiFi-GAN leverages three training loss functions: LS-GAN, STFT Loss and Feature Matching Loss that improve the stability and efficiency of adversarial training, as well as improve the perceptual audio quality.
While the HiFiGAN authors report that they are able to perform mel-spectrogram inversion for unseen speakers, in practice we find that we have to fine-tune the HiFiGAN vocoder on the unseen speaker's speech data to obtain acceptable audio quality.

\subsection{Finetuning Methods}
To adapt a pre-trained single-speaker TTS model for a new speaker, we finetune both the spectrogram-synthesizer and vocoder models. While we started finetuning only the spectrogam-synthesizer, we observed a substantial drop in audio quality when vocoding even real mel-spectrograms of unseen speakers using a pretrained, multi-speaker HiFiGAN vocoder. 
Therefore, we finetune both the the components of the TTS synthesis pipeline using the following approaches:
\subsubsection{Direct Finetuning}

In this approach, we finetune all the parameters of the pre-trained TTS models directly on the data of the new speaker. For finetuning the spectrogram-synthesis model, we require the text and speech pairs of the new speaker, while for the vocoder model, we only require the speech examples of the speaker for generating the spectrogram and waveform pairs. We use mini-batch gradient descent with Adam optimizer~\cite{kingma2014adam} using a fixed learning rate.

\subsubsection{Mixed Finetuning}
Direct finetuning can result in overfitting or catastrophic forgetting~\cite{catastrophic} when the amount of training data of the new speaker is very limited. To address this challenge, we explore another transfer learning method in which we mix the original speaker's data with the new speaker's data during finetuning. 
In this setting, we assume that we have enough training samples of the original speaker while the number of samples of the new speaker is limited. We create a data-loading pipeline that samples equal number of examples from the original and new speaker in each mini-batch. 

Since we now have training data from two speakers we modify the architecture of the FastPitch model and add a speaker embedding layer to make it a two-speaker synthesizer. During training, the we lookup the speaker embedding from the speaker id of each training sample, and add it to the text embedding at each time-step before feeding the input to the first FFTr of the FastPitch. That is,
\begin{equation}
    h= \textit{FFTr}(x + \textit{Repeat}(\textit{speakerEmb}))
\end{equation}

While the FastPitch model parameters are loaded from the pre-trained checkpoint, the speaker embedding layer is randomly initialized and trained along with the other parameters of the model.

Finetuning the vocoder on mixed data is easier than the spectrogram synthesizer. Since the spectrogram representation already contains speaker-specific attributes, we do not require additional speaker conditioning. We simply finetune the model parameters on the spectrogram and waveform pairs from the two speakers. Similar to the spectrogram-synthesizer, we use mini-batches with balanced data from the two speakers during finetuning.

\section{Experiments}

\setlength{\tabcolsep}{6pt}

\begin{table*}[t]
\centering
\resizebox{1.0\textwidth}{!}{%
\begin{tabular}{lr|cc|r|rrr|rr}
\multicolumn{2}{c}{} & \multicolumn{2}{c}{\emph{MOS - Naturalness}} & \multicolumn{1}{c}{\emph{SV}} & \multicolumn{3}{c}{\emph{Pitch Errors (\%)}} & \multicolumn{2}{c}{$\#\textit{phoneme}/\textit{sec.}$}\\
\toprule
Method & Train Data & Speaker 92 & Speaker 6097 & EER & GPE & VDE & FFE & Spk 92 & Spk 6097\\
\midrule
Real Data& - & $3.94 \pm 0.17$ & $3.91 \pm 0.14$ & $9.67$ & - & - & - & $13.1$ & $13.9$ \\
From Scratch    & $>27$ hours & $3.44 \pm 0.15$ & $3.29 \pm 0.14$ & $8.61$ & $28.5$	& $15.5$	& $29.1$ & $13.6$ & $13.7$ \\
\midrule
Direct Finetuning & $1$ min & $2.81 \pm 0.19$  & $2.88 \pm 0.18$ & $10.99$ & $34.9$	& $21.8$ & $37.1$ & $16.3$ & $15.3$\\
Direct Finetuning & $5$ mins & $3.21 \pm 0.15$ & $3.19 \pm 0.15$ & $8.15$ & $28.3$	& $17.0$ & $30.35$ & $15.0$ & $15.1$\\
Direct Finetuning & $30$ mins & $3.40 \pm 0.16$  & $3.41 \pm 0.14$ & $8.69$ & $28.2$ & $19.9$ & $32.4$ & $14.6$ & $14.9$ \\
Direct Finetuning & $60$ mins & $3.39 \pm 0.16$  & $3.31 \pm 0.14$ & $9.48$ & $27.1$ &	$18.5$	& $30.3$ & $14.6$ & $14.7$\\
\midrule
Mixed Finetuning & $1$ min & $2.82 \pm 0.20$   & $2.55 \pm 0.17$ & $11.12$ & $33.3$ & $23.6$ & $37.0$ & $16.8$ & $16.4$\\
Mixed Finetuning & $5$ mins & $3.33 \pm 0.16$  & $3.13 \pm 0.15$ & $9.51$ & $29.3$	& $19.8$ & $32.0$ & $15.6$ & $14.8$\\
Mixed Finetuning & $30$ mins & $3.21 \pm 0.17$ & $3.22 \pm 0.14$ & $9.68$ & $30.8$	& $17.3$ & $31.3$ & $14.1$ & $14.9$\\
Mixed Finetuning & $60$ mins & $3.32 \pm 0.16$ & $3.19 \pm 0.14$ & $9.66$ & $32.2$	& $22.2$ &	$34.4$ & $13.9$ & $14.8$\\
\bottomrule
\\
\end{tabular} 
}
\caption{Naturalness, speaker verification and style similarity evaluations for different finetuning methods and benchmarks. We also present results on real validation data and a models trained from scratch on more than 27 hours of data for each speaker. }
\label{tab:naturalness}
\end{table*}

We focus our experiments on the publicly available Hi-Fi TTS dataset~\cite{bakhturina2021hi}. The Hi-Fi TTS dataset contains about 292 hours of speech from 10 speakers with at least 17 hours per speaker sampled at 44.1 kHz. 
We train a single-speaker TTS model on speaker 8051 (Female) from the dataset and perform finetuning experiments on speaker 6097 (Male) and speaker 92 (Female). 
Each of these three speakers contain at least 27 hours of data. We keep aside 50 text and speech pairs from each speaker as validation samples. 

We create 4 training subsets for each target speaker (92 or 6097) of varying sizes: 1 minute, 5 minutes, 30 minutes and 60 minutes. We do this for both speaker 92 and 6097. 
For the \textit{mixed finetuning} approach, we mix the new speaker's data with 5000 samples ($\sim5$ hours) from the original speaker 8051.
We perform mini-batch gradient descent for ($200 \times \textit{data-size in minutes}$) iterations for the direct finetuning method and  ($1000 \times \textit{data-size in minutes}$) iterations for the mixed finetuning method. We use fewer iterations for the direct finetuning method to prevent overfitting. On a single Tesla-V100 GPU, direct finetuning on 5 minutes of data takes less an hour of training time. 

We synthesize speech for the text samples in the validation set of each speaker and evaluate the synthesized speech on three aspects --- naturalness, voice similarity to the target speaker and speaking style similarity to the target speaker. To perform these evaluations, we synthesize speech samples for the unseen text samples in the validation set of the target speaker. 
We also present audio examples and a finetuning demo to adapt pretrained TTS models on a small dataset~\footnote{Audio Examples and Demo: \href{https://paarthneekhara.github.io/tlfortts/}{https://paarthneekhara.github.io/tlfortts/}}.

\subsection{Naturalness}
To assess speech naturalness, we conducted a crowd-sourced listening test on Amazon Mechanical Turk (AMT) and asked listeners to rate each audio utterance on a 5-point naturalness scale (1 to 5 with 0.5 intervals) to collect Mean Opinion Scores (MOS). Each sample of the validation set (synthetic or real) is rated by 4 independent listeners resulting in 200 evaluations of each technique per speaker. 
We present the naturalness MOS scores with 95\% confidence intervals in Table~\ref{tab:naturalness}. 
We notice that finetuning on $\geq 5 \text{minutes}$ of data can achieve similar MOS scores as compared to a model trained from scratch on more than 27 hours of data. 
Interestingly, both mixed and direct fine-tuning methods achieve similar MOS scores. However, we observe that the mixed finetuning method performs slightly better than the direct finetuning approach in pronunciations of unseen words. We encourage the readers to listen to the audio examples linked in the footnote to observe this difference. 
\begin{figure}[t]
    \centering
    \includegraphics[width=1.0\linewidth]{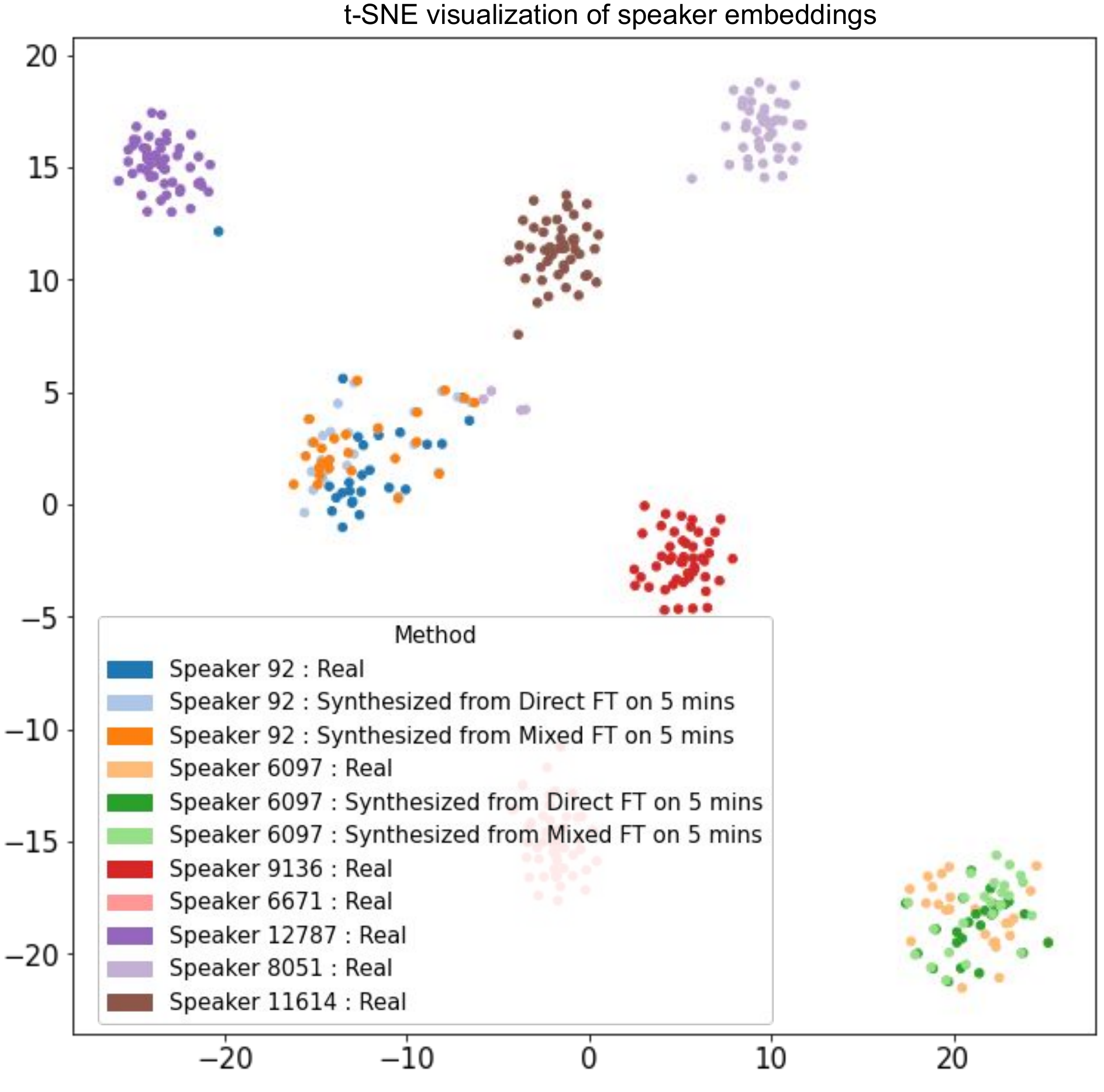}
    \caption{
    Speaker embedding visualization of real and synthetic utterances. For a given speaker, the synthetic embeddings (eg. light blue and orange for speaker 92) are closely clustered with the real utterance embeddings (eg. dark blue for speaker 92) and well separated from other speakers.
    }
    \label{fig:tsne}
\end{figure}

\subsection{Voice Similarity}

To evaluate the voice similarity of the synthesized speech with our target speaker, we perform speaker verification evaluations using a pre-trained speaker verification model~\cite{Koluguri2020SpeakerNet1D}. 
The speaker verification model is a depth-wise separable convolutional network that is trained on the VoxCeleb1~\cite{nagrani2017voxceleb} and VoxCeleb2~\cite{chung2018voxceleb2} dev sets. 
We first visualize the speaker embeddings of real and synthetic data by reducing the 256 dimensional utterance embeddings to 2 dimensions using t-SNE. We find that the speaker embeddings of synthetic audio of a given speaker are very closely clustered with the embeddings of real validation samples of the same speaker. 
We present the embedding visualization of real data and audio synthesized from finetuning using 5 minutes of data in Figure~\ref{fig:tsne}.

To quantitatively evaluate the speaker clustering, we perform Equal Error Rate (EER) evaluations that are commonly used for evaluating speaker verification systems.
We first evaluate the speaker verification EER on trial pairs from the actual validation data of Hi-Fi TTS --- We pair each validation example from our target speaker (92 or 6097) with all the other validation examples of the same speaker to create positive pairs and then with validation examples of alternate speakers (all other speakers in the Hi-Fi TTS dataset) to create negative pairs. This results in a total of $49,900$ trials.



Next, we synthesize speech samples from the validation text examples of speaker 92 and 6097 and pair them up with the same speaker real's validation audio to create positive pairs and a different speaker's validation audio to create negative pairs resulting in the same number of trials as the real data evaluation. 
We compute the Equal Error Rate (EER) metric on these trial pairs and report the results in Table~\ref{tab:naturalness}. We find that the EER on synthetic data is very close to that on real data which is in agreement with our observation from the t-SNE plots.



\subsection{Style similarity}

\label{sec:feature}

While the low speaker verification EER indicates that our finetuning methods can closely mimic the timbre of the target speaker, we also need to evaluate other style aspects of synthesized speech. Following past works~\cite{mellotron,neekhara2021expressive} we evaluate the error metrics for the pitch (fundamental frequency) contours of the synthetic and real validation samples. We calculate Gross Pitch Error(GPE)~\cite{nakatani2008method}, Voicing Decision Error(VDE)~\cite{nakatani2008method} and F0 Frame Error~\cite{chu2009reducing}. In order to generate same length audio samples as the real data, we use forced alignment to extract the phoneme duration during inference. We present the results in Table~\ref{tab:naturalness}. 

We also calculate the speaking rate characteristics in terms of phonemes per seconds on the ground truth validation samples and the speech synthesized directly from text without using forced alignment. We find that increasing the amount of training data reduces the difference between the speaking rate of actual data and synthetic speech. For both speakers, we observe that the speaking rate of synthetic speech is much faster than that of the actual data when we use $\leq5$ minutes of data. We recommend future work to investigate this issue of faster speaking rate when we use limited data for finetuning.

\section{Conclusion}
In this work, we studied transfer learning methods for adapting TTS models for a new voice. 
We demonstrate that pre-training the TTS model on just one speaker can be very effective in reducing the training data and compute required for generating a new voice. We find that using just 30 minutes of transcribed audio of the new speaker, our methods can yield competitive results in terms of naturalness, voice similarity and style similarity as compared to a model trained from scratch on more than 27 hours of data. 
We open-source our finetuning methods that enable training of high-quality TTS models in limited data and resource constrained settings.
We recommend future work on developing a reliable validation metric based on which we can determine the optimal number of finetuning iterations.

\bibliographystyle{IEEEtran}

\bibliography{mybib}


\end{document}